\documentclass[12pt,preprint]{aastex}
 
\newcommand{\degK}{~^{\mbox{o}}\mbox{K}}

\newcommand{\dyn}{\mbox{~dyn}}
\newcommand{\cm}{\mbox{~cm}}

\newcommand{\kms}{\mbox{~km s}^{-1}}
\newcommand{\kpc}{\mbox{~kpc}}

\newcommand{\Myr}{\mbox{~Myr}}

\newcommand{\twoarm}{(1, 0.175, 2)}
\newcommand{\nonmag}{(2.5, 0, 2)}
\newcommand{\fourarm}{(1, 0.175, 4)}
  
\begin{document}

\title{3D MHD Modeling of the Gaseous Structure of the Galaxy:
       Setup and Initial Results.}

\author{Gilberto C. G\'omez}
\affil{Department of Astronomy, University of Wisconsin-Madison,
  475 N. Charter St., Madison, WI 53706 USA}
\email{gomez@wisp.physics.wisc.edu}
\and
\author{Donald P. Cox}
\affil{Department of Physics, University of Wisconsin-Madison,
  1150 University Ave., Madison, WI 53706 USA}
\email{cox@wisp.physics.wisc.edu}

\begin{abstract}

We show the initial results of our 3D MHD simulations of the
flow of the Galactic atmosphere as it responds to a spiral
perturbation in the potential.
In our standard case, as the gas approaches the arm, there is a
downward converging flow that terminates in a complex of
shocks just ahead of the midplane density peak.
The density maximum slants forward at high $z$, preceeded by
a similarly leaning shock.
The latter diverts the flow upward and over the arm,
as in a hydraulic jump.
Behind the gaseous arm, the flow falls again, generating further
secondary shocks as it approaches the lower $z$ material.
In cases with two arms in the perturbing potential, the gaseous
arms tended to lie somewhat downstream of the potential
minimum. In the four arm case, this is true at large $r$ or
early evolution times.
At smaller $r$, the gaseous arms follow a tighter spiral,
crossing the potential maximum, and fragmenting into sections
arranged on average to follow the potential spiral.
Structures similar to the
high $z$ part of the gaseous arms are found in the
interarm region of our two-armed case,
while broken arms and low column density
bridges are present in the four-armed case.
Greater structure is expected when we include cooling of denser
regions.

We present three examples of what can be learned from these models.
We compared the velocity field with that of purely
circular rotation, and found that an observer inside the galaxy
should see radial velocity deviations typically greater than
$20 \kms$.
Synthetic spectra, vertical from the midplane,
show features at velocities
$\approx -20 \kms$, which do not correspond to
actual density concentrations.
Placing the simulated observer outside the galaxy,
we found velocity structure and arm corrugation
similar to those observed in H$\alpha$ in NGC 5427.

\end{abstract}


\section{INTRODUCTION.}

Even though spiral structure is one of the most prominent
features of disk galaxies, details of the spiral
arms in our own Galaxy remain uncertain.
\citet{geo76} traced the spiral structure of the Milky Way using
\ion{H}{2} regions, and developed a model with four arms.
More recent attempts concluded that the Milky Way might actually
have a superposition of two and four arm structures, each
one with different pitch angles, which might
arise from different components of the galactic disk
\citep{dri00, lep01}, suggesting that the stellar and gaseous
disks might not be tightly coupled.
Similar behavior has been frequently observed in external
galaxies \citep[for example]{pue92, gro98}.

\citet{rob69} showed that the gas must generate a large scale shock
in the presence of a spiral perturbation.
It was proposed that
the density enhancement induced by this shock might generate
a sequence of molecular clouds and star formation downstream
from the shock, which itself was associated with the strong
dust lane observed in the inner region of the spiral arms
in external galaxies.
Two dimensional numerical models by \citet{tub80} and \citet{sou81}
showed that the gas forms a vertical shock perpendicular
to the plane of the galactic disk.
The post-shock gas
remained close to hydrostatic equilibrium, even with an adiabatic
equation of state.
Their results did not show vertical motions larger than $3 \kms$.
In fact, the largest downflow they found was due to the pre-shock
gas readjusting its vertical structure as it flows into the arm
potential.
Therefore, when \ion{H}{1} observations on face-on galaxies showed
extended velocity components with dispersions of the order
of $20 \kms$, they were attributed to other phenomena, such as
galactic fountains, a warping of the \ion{H}{1} disk, or
intermediate velocity clouds \citep{dic90, kam92, kam93}.

Since then, we have realized that the ISM is thicker and with a
higher pressure than previously thought.
The pressure scale height has been found to be larger than
the density scale height, and the non-thermal pressures
(turbulent, magnetic and cosmic ray) are at least as large as
the thermal component
\citep{bad77, rey89, bou90}.
Therefore, less compressible gas needs to be considered in order
to generate more realistic models of the ISM.
Such a medium, with a larger effective
$\gamma$ (the ratio of the specific
heats) would be more likely to display the vertical motions
characteristic of a hydraulic jump.
With this in mind,
\citet[MC]{mar98} performed 2D MHD simulations of
the flow of the gaseous disk
and found diverse structures that differed from
the vertical near-hydrostatics found in previous studies.
In many cases, the gas moved up ahead of the stellar arm,
sped up over it, and fell behind with large bulk velocity.
Frequently, there was a downstream shock at higher $z$
as this downflow was arrested, sometimes resulting in secondary
midplane density maxima.

The goal of our investigation is to extend calculations
like those of MC to three dimensions,
to a large fraction of the Galaxy,
and to look for its possible observational signatures.
In this paper, we present the early results of these simulations.
In Section 2 we describe the numerical setup and the procedure
to achieve the initial hydrostatic equilibrium,
in Section 3 we describe the results of the simulations,
in Section 4 we present three examples of synthetic observations
that can be done with this type of simulation, and in Section
5 we present our conclusions.

\section{THE NUMERICAL SETUP.}

We performed 3D MHD simulations in polar coordinates using the
code ZEUS \citep{sto92a, sto92b, sto92c}.
This code solves the ideal MHD equations for an inviscid
fluid with infinite conductivity in a fixed eulerian grid.
For our standard case, the grid extends from
0 to 1 kpc in $z$, 3 to 11 kpc in $r$, and 0 to $2 \pi/N$ in the
azimuthal angle $\phi$ with 50, 80 and $200/N$
grid points in each respective direction, $N$ being the number
of arms in each case.
The boundaries in $r$ and $z$ are reflective, while those
in $\phi$ are periodic.

\subsection{Hydrostatics, Theory.}

The azimuthally averaged gravitational potential used
is model 2 from \citet{deh98}.
Using this potential, we set up a hydrostatic interstellar medium
based on the scheme introduced by \citet{yor82} and described
in the appendix of \citet{ben02}.
We extended the procedure to include the effects of
an azimuthal magnetic field whose magnitude depends only on
the density.
Given a density profile in the midplane, an equation of
state (isothermal, in our case)
and a density-magnetic pressure relation,
the density and rotation velocity are uniquely defined everywhere.

Define the function $G(\rho)$ as:

\begin{equation}
\label{eq1}
G(\rho)=\int \frac{d}{d\rho}(p_T+p_B) \frac{d\rho}{\rho},
\end{equation}

\noindent
where $\rho$ is the density and $p_T$ and $p_B$ are
the thermal and magnetic pressures, respectively.
Vertical hydrostatics,
$\partial(p_T+p_B)/\partial z = - \rho \partial \Phi/\partial z$
where $\Phi$ is the gravitational potential,
reduces via Equation \ref{eq1} to:

\begin{eqnarray}
\frac{\partial (G + \Phi)}{\partial z} &=& 0 \label{eq2} \\
\Rightarrow G[\rho(r,z)] &=& G[\rho(r,0)]+\Phi(r,0)-\Phi(r,z).
\label{eqA}
\end{eqnarray}

\noindent
At any $z$, the velocity profile is given by
the radial balance between the
radial potential and pressure gradients, the magnetic tension and
the centrifugal force:

\begin{equation}
\frac{v_\phi^2(r,z)}{r}
 = \frac{\partial \Phi}{\partial r}
  +\frac{1}{\rho(r,z)}\left[\frac{\partial}{\partial r}(p_T+p_B)
  +\frac{2 p_B(r,z)}{r}\right].
\label{eqC}
\end{equation}

\noindent
This balance can be reduced to:

\begin{equation}
\frac{\partial (G+\phi)}{\partial r}
= \frac{v_\phi^2 - 2 p_B/\rho}{r}
= \frac{v_\phi^2 - v_A^2}{r},
\label{eq5}
\end{equation}

\noindent
where $v_A$ is the Alfv\'en velocity.
Provided that $\partial^2(G+\Phi)/\partial z \partial r = 
               \partial^2(G+\Phi)/\partial r \partial z$,
along with Equations \ref{eq2} and \ref{eq5}, we have:

\begin{eqnarray}
\frac{\partial (v_\phi^2 - v_A^2)}{\partial z} &=& 0 \\
\Rightarrow v_\phi^2(r,z) &=& v_\phi^2(r,0) -v_A^2(r,0) + v_A^2(r,z).
\label{eqB}
\end{eqnarray}

\noindent
Given the midplane density distribution, $\rho(r,0)$, the
density off the plane is obtained by solving Equation \ref{eqA}
for $\rho(r,z)$, while $v_\phi(r,0)$ is obtained from Equation
\ref{eqC} evaluated at $z=0$, after which the equilibrium
rotation speed at all other $z$ follows from Equation \ref{eqB}.

\subsection{Hydrostatics, Implementation.}

\citet{mar98} performed their 2D MHD  calculations of spiral
arm structure using the vertical density distribution at the
solar circle compiled by \citet{bou90}, modified to have a
slightly lower vertical scale height for the warm ionized
component.
We found that the vertical distribution can be
reproduced fairly accurately with thermal and magnetic
pressures as follows.
The thermal component assumes a neutral gas with a constant
temperature of $5700 \degK$ and an isothermal equation of state.
The magnetic pressure is taken as:

\begin{equation}
p_B=1.75 \times 10^{-12} \frac{n}{n+n_c} \dyn\cm^{-2},
\label{eq8}
\end{equation}

\noindent
where $n_c=0.04 \cm^{-3}$.
The form of the magnetic pressure is such that it has little
gradient at high density, and is proportional to $\rho$ at
low density.
The former accomodates a dense thermally supported core near
the midplane, while the latter leads to a higher but constant
signal speed at low density, far off the plane.

With a helium abundance equal to 10\% of the hydrogen
abundance by number, the mean atomic mass is $(14/11)m_H$.

In our initial work, we wanted to explore a situation that was
not so heavily dominated by magnetic pressure.
We therefore raised the temperature to $10^4 \degK$ and reduced
the magnetic pressure by a factor of 10, keeping 
$n_c=0.04 \cm^{-3}$.
The midplane density distribution was taken as exponential,
with a radial scale length of 4 kpc and a density of
$1.11 \cm^{-3}$ at $r=8 \kpc$.

\begin{figure}
\plotone{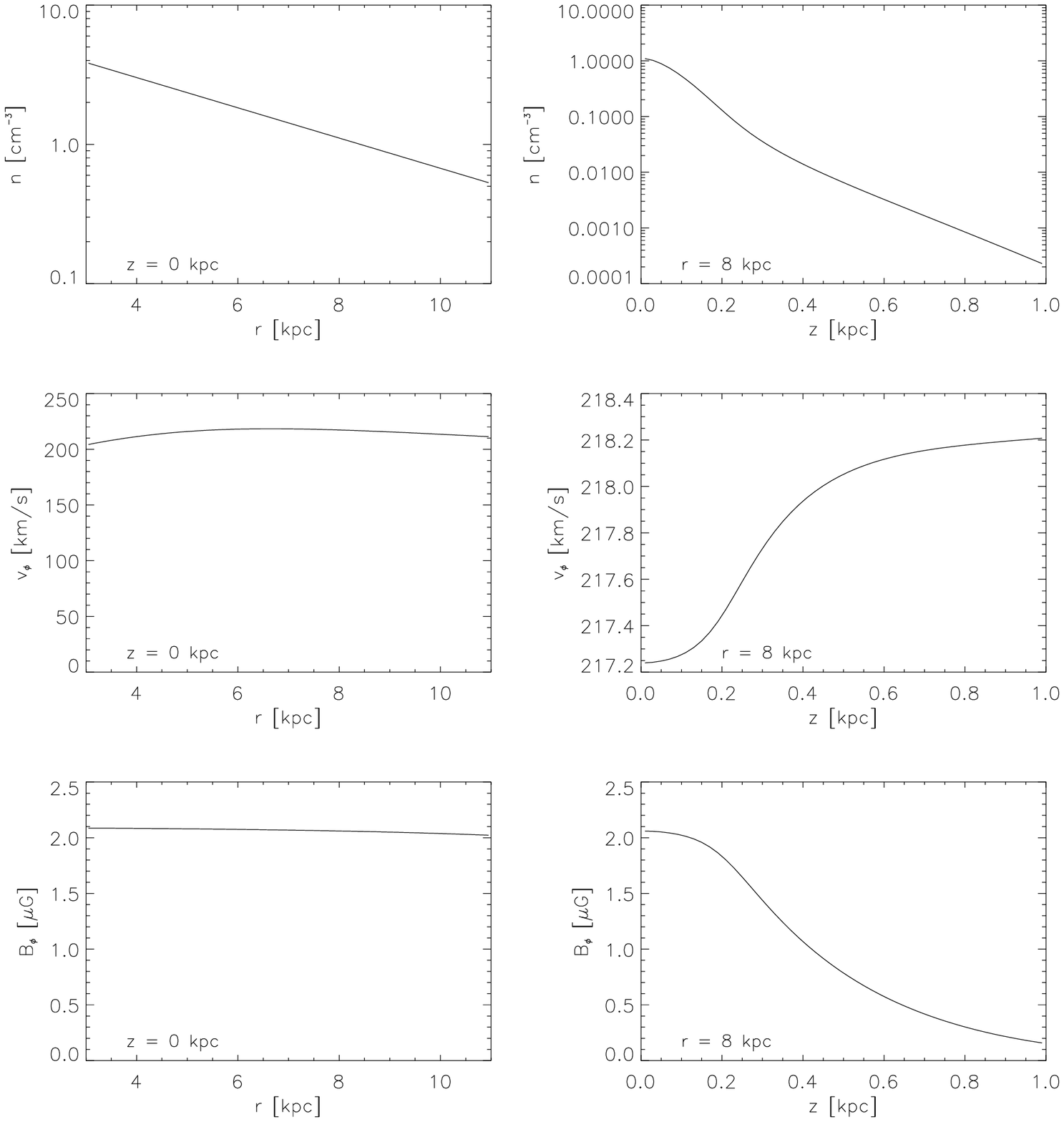}
\caption[]{
Initial state of the \twoarm~ case.
Cases are labeled with ($T/10^4 \degK,
~p_{mag}/ 10^{-12} \dyn \cm^{-2}, N$),
where $T$ is the temperature, $p_{mag}$ is the coefficient in
Equation \ref{eq8}, and $N$ is the number of spiral arms.
The gas is in hydrostatic and dynamical equilibrium in the
vertical and radial directions with a temperature of $10^4 \degK.$
The density is presented assuming a mean
particle mass $\mu = 1.27 m_H$.

}
\label{fig1}
\end{figure}

When these parameters are introduced into the above formalism
and the hydrostatics found, the vertical half disk
column density at $r=8 \kpc$ is $0.12 \kpc \cm^{-3}$.
Figure \ref{fig1} shows the density, rotation velocity and
magnetic field strength versus radius at the midplane, and
versus $z$ at $r=8 \kpc$.
The midplane density varies by less than a factor of 10,
while vertically the density drops nearly four orders of
magnitude between the midplane and $z=1 \kpc$, our present
maximum height.
At smaller radii, the vertical gravity is stronger
and the density gradient in $z$ even larger, so that both the
highest and lowest densities occur at the inner boundary.
A curious feature is that at high $z$, the density increases
with increasing radius: the disk ``flares''.

The rotation velocity varies only slightly with radius,
by less than $15 \kms$, and by much less with $z$,
only about $1 \kms$ increase (as per Equation \ref{eqB},
higher Alfv\'en speed requires higher rotation rate at high $z$
in this approximation).
The magnetic field strength varies only slowly with radius, and
appears roughly Gaussian in $z$, with a flat region in
the inner 200 pc of the thermal core.

We also report below on a case which has no magnetic
field, in which the constant value of the temperature
was taken as $2.5 \times 10^4 \degK$.
More precisely, the thermal pressure at a given density
was taken as 2.5 times that of the previous run, because the
above temperature was used inconsistently with assuming the
gas was still neutral.
This increase in thermal pressure was made in order to have
a density distribution roughly similar to the magnetic case.

\subsection{The Spiral Perturbation.}

In addition to the axisymmetric potential,
we used a spiral perturbation of fixed
shape that rotates with $\Omega_P = 12 \kms \kpc^{-1}$ and has
a pitch angle of $15 \deg$.
Details are reported in \citet{cox02}.
All the simulation grid is inside corotation.
The depth of the perturbation varies slightly in $r$, weakens
in $z$ and has a sinusoidal profile in $\phi$;
in the midplane, its corresponding mass
density amplitude is $\approx 52\%$ of the
disk component of the axisymmetric model at $r=8 \kpc$,
which provides a peak to valley potential
difference  of about $(30 \kms)^2$.
This mass contrast is consistent with
{\it K}-band observations performed by \citet{rix93}, \citet{rix95}
and \citet{kra01},
who quote an arm/interarm contrast in density of old stars
between 1.8 and 3 for a sample
of spiral galaxies.

\subsection{Numerical Complications.}

In early runs, performed with outflow boundary conditions
at the inner, outer and upper boundaries, ZEUS soon reported
difficulties with ``hot zones'' in which the timestep became
so short that the calculation terminated.
This is almost certainly caused by the enormous density
contrast of our hydrostatic solution.
By making those three boundaries reflecting, so that the material
is unable to flow off the grid, this problem was postponed
or eliminated, depending on the case run.
The two arm magnetized case ran to about 270 Myr, the unmagnetized
case and the four arm magnetized case ran
the full 400 Myr asked of them.

In addition to changing the boundary conditions,
we also changed the perturbation force field near the inner
boundary after noticing that the spiral potential was pushing
material against the inner boundary, causing reflected
waves that propagated outward.
In order to avoid splashing against the inner-$r$ boundary, this
perturbation is not applied in the inner 1 kpc of the grid, while
it is smoothly turned on in the subsequent kpc.
Thus, the useful computational grid runs from 5 to 11 kpc.
Also, in order to diminish initial transient effects,
the spiral perturbation is turned on gradually
during the first 50 Myr.
This short turn-on time undoubtedly creates part of the transient
behavior and may have exaggerated some of the early velocity
structure.
Our intention is to make runs lasting so long that such transients
have died out, and to report on that asymptotic behavior in future
work.

We are wary of artifacts that might be caused by our boundary
conditions and will continue to experiment with alternatives,
including cases with lower overall density contrast that
might allow open boundaries.

\section{BEHAVIOR OF THE SIMULATIONS.}

Our primary example is the two armed spiral with moderate
(reduced from MC) magnetic pressure.
Our calculation space was the upper half (in $z$) of half
a circular disk, with periodic boundaries in $\phi$.
The period of time for a mass element to rotate around
this half disk, relative to the rotating pattern, is about
100, 200 and 340 Myr ar $r=5, 8$ and 10 kpc,
respectively.
We have chosen a fiducial time of 248 Myr, for our initial
examination of the structure.
At this time, nearly all mass elements have experienced the spiral
perturbation once or twice,
but conditions are still transient, representative
of local interaction with rather than global accomodation to the
perturbation.
We will compare this early structure with that of
the unmagnetized case at the same time, and the 4 arm magnetized case at half that time, which roughly represents the same
level of maturity.

Having examined those single early time characteristics,
we will present features of the subsequent development of the
2 arm cases and the much more mature 4 arm case,
as indicative of features requiring a longer time to appear.

In the remainder of this paper, we will refer to each case
using a three element naming scheme, describing the
temperature (in units of $10^4 \degK$), the numerical
coefficient in the magnetic pressure-density relation
(Equation \ref{eq8}, in units of $10^{-12} \dyn \cm^{-2}$)
and the number of spiral arms in the perturbation potential.
Therefore, the magnetic two arm case will be denoted \twoarm,
the non-magnetic two arm case \nonmag, and the
four arm magnetic case \fourarm.

\subsection{Two Arm Magnetized Case.}

\begin{figure}
\plotone{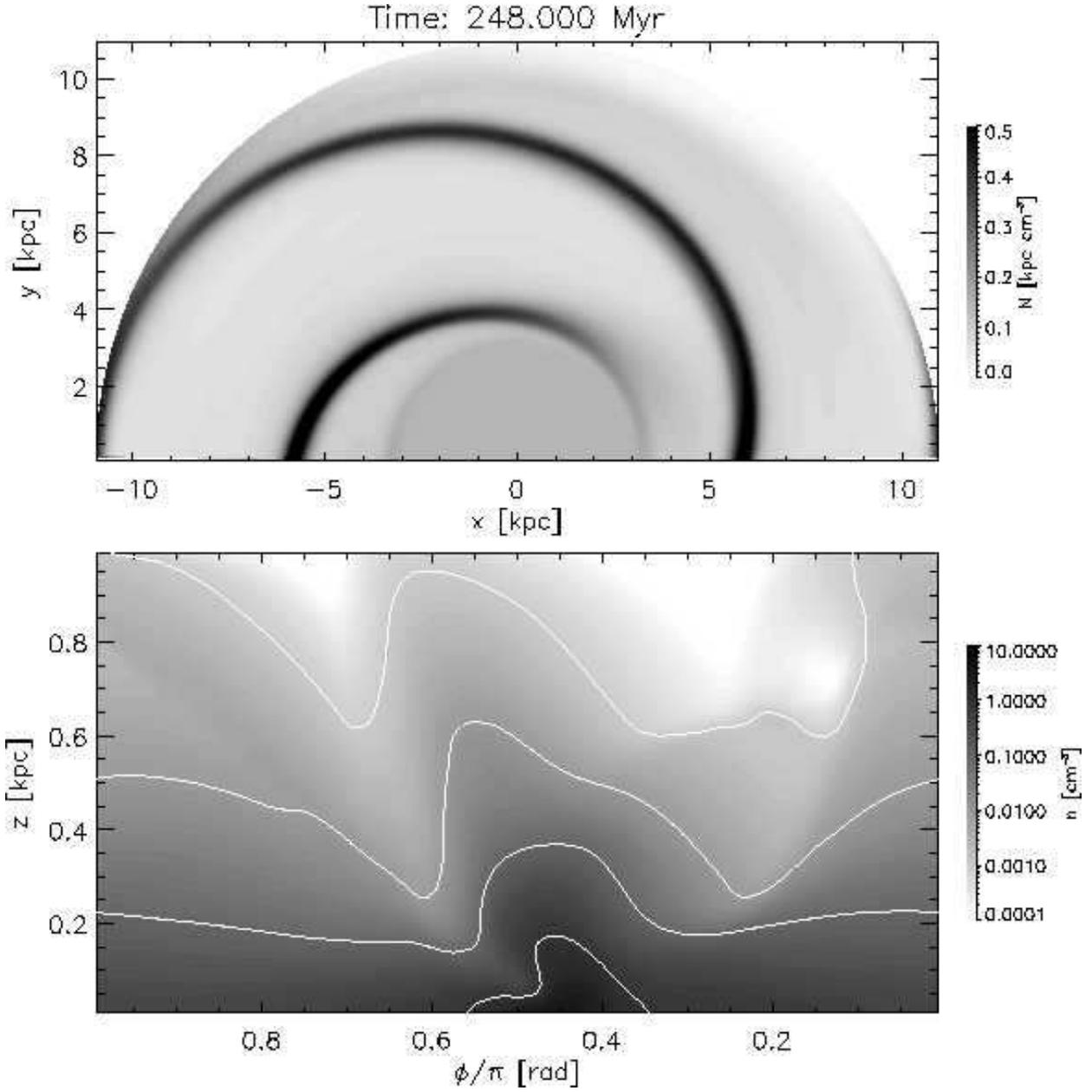}
\caption[]{
The upper panel shows the column density of the simulation for
the \twoarm~ case.
The lower panel shows density along a cylindrical surface
at $r=8 \kpc$.
White contours show density increasing in factors of 10, starting
at $10^{-3} \cm^{-3}$.
}
\label{fig2}
\end{figure}

\begin{figure}
\plotone{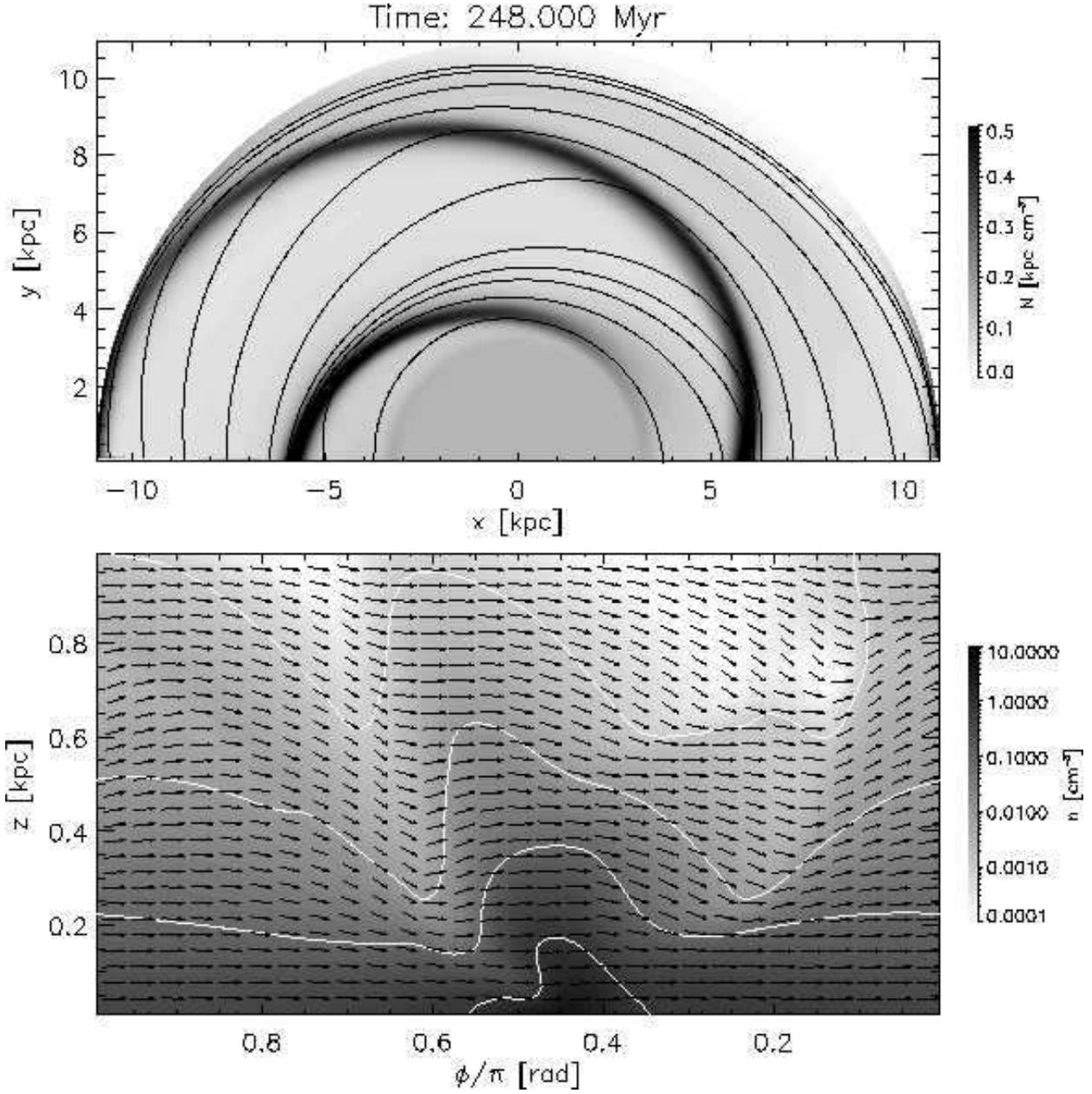}
\caption[]{
Same as Figure \ref{fig2}, with velocity structures superimposed.
The continuous lines in the upper panel
show the clockwise velocity field of the gas
in the midplane.
In the lower panel, the direction
and relative length of the arrows is correct, but the length
of the individual components is not. See discussion in the text.
}
\label{fig3}
\end{figure}

Figures \ref{fig2} and \ref{fig3} show the results
for the \twoarm~ case.
The lower panel shows the density and (in Figure \ref{fig3})
velocity field
along a surface of constant radius $r=8 \kpc$.
The upper panel shows column density for $z \ge 0$, half
the total.
In the upper panel, rotation is clockwise, and in Figure \ref{fig3},
the lines show the integrated velocity field at the midplane.
For clarity in the visualization, we modified the components
of the velocity in the lower panel of this Figure, so the arrows
representing velocity in the inertial reference frame are
parellel to the flowlines, with relative lengths proportional
to the total velocity.

\begin{figure}
\plotone{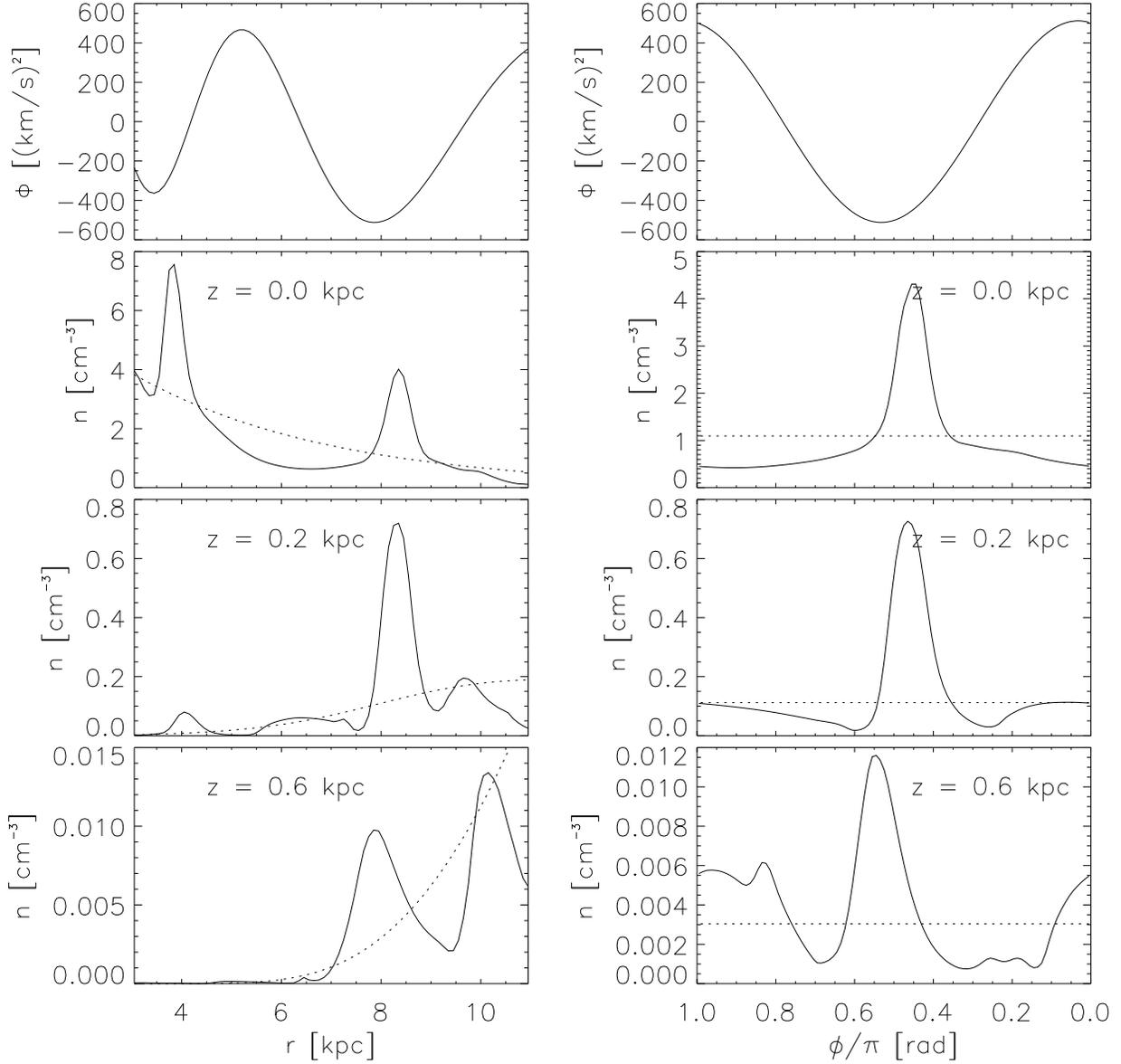}
\caption[]{
Midplane perturbation potential, and
density at different heights for the \twoarm~ case at
248 Myr,
along $\phi=\pi/2$ and $r=8 \kpc$.
Dotted lines show the initial (hydrostatic) density.
}
\label{fig4}
\end{figure}

The most important feature in Figure \ref{fig2} and \ref{fig3}
is the presence of a simple grand design spiral.
The density concentration in the midplane contains most of the
column density. At higher $z$, this feature
leans forward.
The midplane gaseous arm appears slightly after the perturbation
potential minimum (outward in radius),
shifting to a better alignment farther above
the plane (Figure \ref{fig4}).
Note also the strong arm to interarm contrast; a significant
fraction of the material is located in the arms.
The convergences of the
velocity flow field in the upper panel of Figure \ref{fig3}
are consistent with this concentration.

\begin{figure}
\plotone{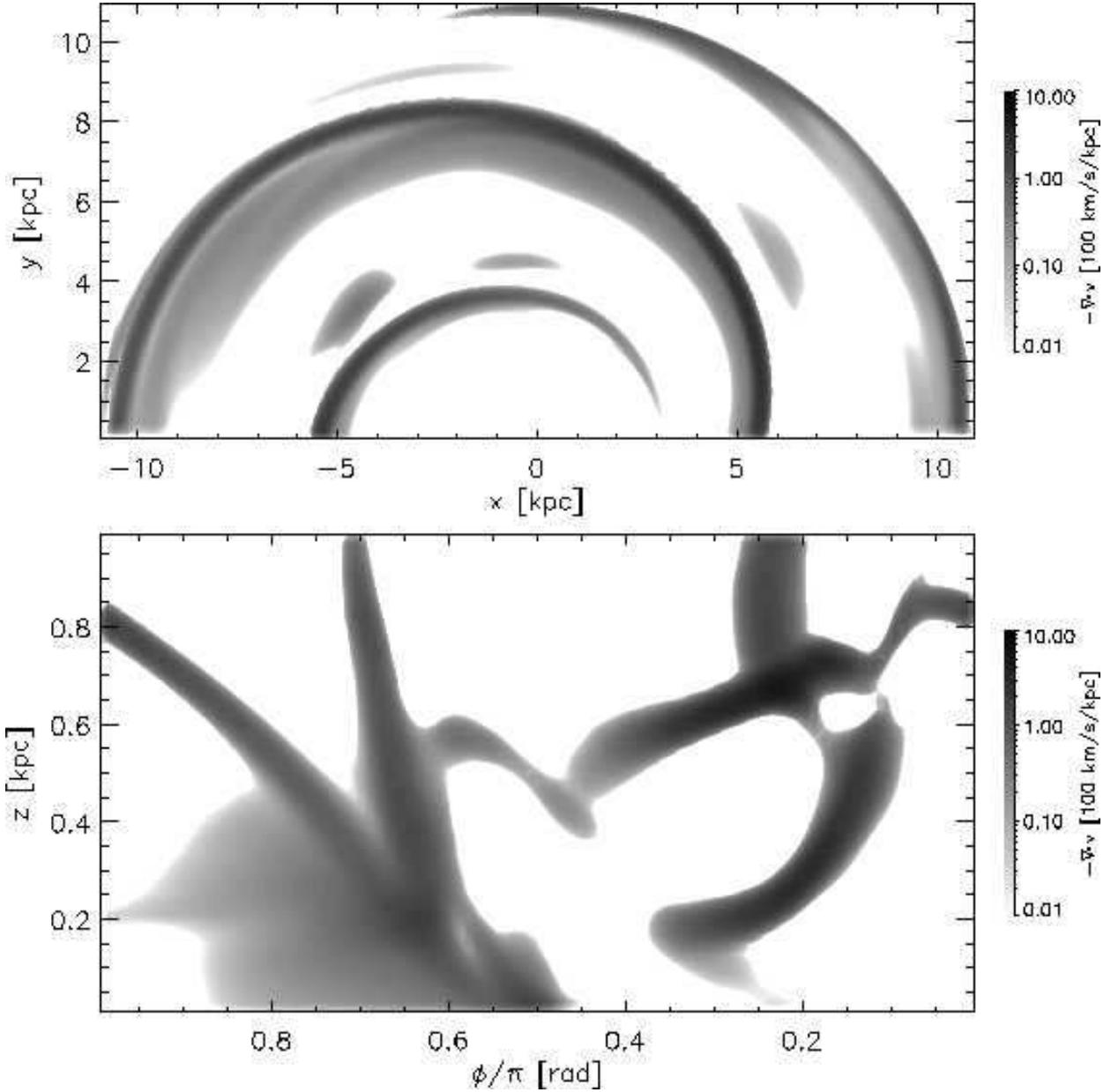}
\caption[]{
Divergence of the velocity for \twoarm~ case at 248 Myr.
Only those places with $\nabla \cdot \vec{v} < 0$ are shown.
The upper panel shows the midplane distribution, peaking at the
inner edges of the column density arms of Figure \ref{fig2}.
There is a substantial forward leaning shock (between
$\phi/\pi = 0.6$ and $0.7$) preceeding the forward leaning
density ridge
of Figure \ref{fig2}.
(The high density at the base of the latter dominates the column
density maps.) But there are a variety of other shocks as well.
}
\label{fig9}
\end{figure}

Figure \ref{fig9} shows the negative of the velocity divergence
at the same positions as Figure \ref{fig3}. Only the regions
with negative divergence are shown, in order to mark the
places where strong compression (and shocks) appear.
In the midplane, the compression of the material into the
dense features is associated with a complex of intersecting
shocks, following a diffuse compression of the material falling
toward the arm.
Comparison of Figures \ref{fig3} and \ref{fig9} shows
an important shock 200 pc off the plane, preceding
the  gaseous arm, also slanting upstream at higher $z$.
This shock accelerates gas upwards
around $\phi/\pi \approx 0.6$. 
Over the arm, the flow is nearly horizontal.
Behind the arm, it falls
with vertical velocities of the order of $20 \kms$,
forming secondary shocks.
This behavior is more evident along the $\phi$-direction, but
it is also visible in radial plots (not shown),
since the presence of the stellar
arms also induces velocities in that direction.
Motions like these are similar to hydraulic
jumps and were observed in the simulations performed by MC.
MC also found midplane gas concentrations which
they attributed to downstream bouncing of the flow.
In our calculations, such concentrations are (so far) much smaller
or absent in the midplane, but do appear in the lower density
high $z$ gas,
almost as if it tries to form another gaseous arm
between the stellar arms.
This interarm structure has little column density,
but it is evident in the bottom panels of
Figures \ref{fig2} and \ref{fig3},
and the right hand panels of Figure \ref{fig4}.

\begin{figure}
\epsscale{.7}
\plotone{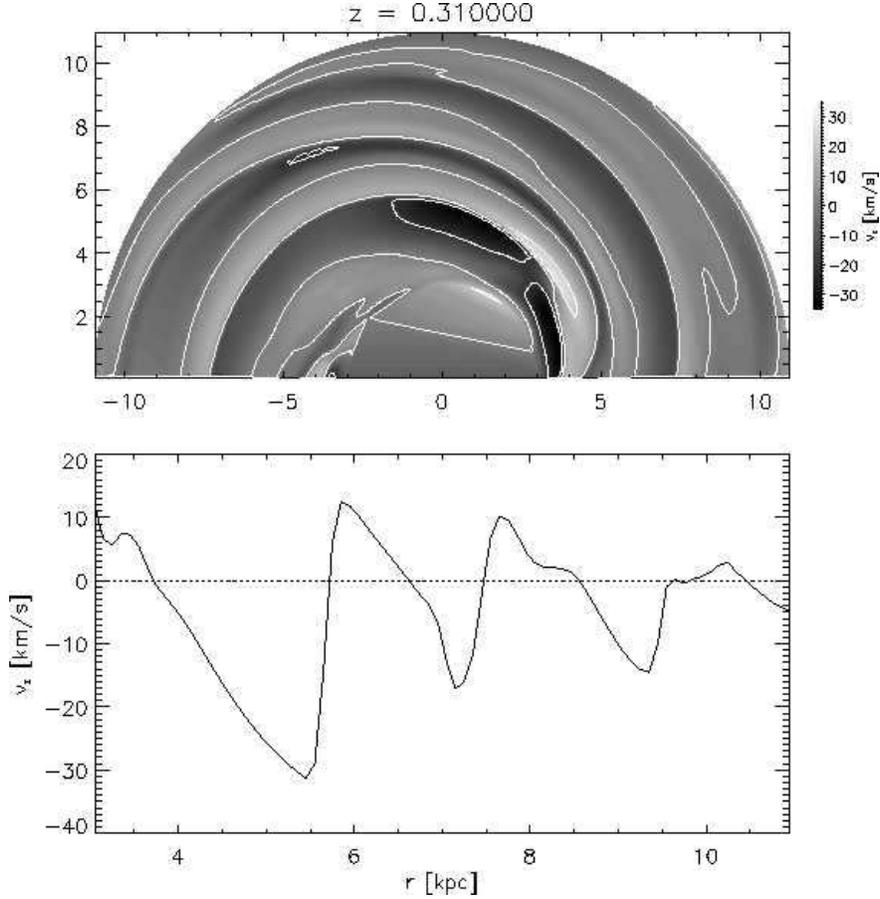}
\caption[]{
Vertical velocity structure at $z=.31 \kpc$ for the \twoarm~ case
at 248 Myr.
White contours in
the upper panel are drawn at $v_z = -20, 0 \mbox{~and~} 20\kms$.
Lower panel shows a cut along the vertical ($\phi=\pi/2$) line.
Comparison of the upper panel with that of Figure \ref{fig2}
shows that the velocity field at this height has twice the frequency
of the column density, i. e., it looks like a four arm spiral.
The sawtooth pattern of the lower panel is consisntent with the
velocity field of Figure \ref{fig3} and $\nabla \cdot \vec v$
in Figure \ref{fig9}.
There is a forward leaning shock (in $z$, see Figure \ref{fig9})
preceeding the forward leaning density ridge of the gaseous arm.
Upstream from that shock, the gas is falling, downstream it is
rising as in a hydraulic jump.
Immediately over the arm, the vertical velocity is close to
zero; gas flows up and over the density ridge and down the other
side.
As found by \citet{mar98}, the falling gas on the
downstream side again shocks.
In their case, this led to secondary density maxima in the
midplane.
In our runs thus far (e. g. Figure \ref{fig4}) there are secondary
density maxima at high $z$, but they are too weak to show
up in column density maps.
}
\label{fig8}
\end{figure}

In Figure \ref{fig8} we present the vertical velocity structure
at 310 pc above the plane.
The upper panel shows a gray scale of $v_z$ with contours at
0 and $\pm 20 \kms$.
The gas moves up and over the gaseous structures at the
arm and at the interarm positions, generating twice as many
spirals in the upper panel of this Figure.
Along the $\phi/\pi=0.5$ direction (lower panel),
this behavior is clearer.
While the arm is at $r \approx 8 \kpc$ in this direction,
the vertical velocity behaves similarly at 6 and 10 kpc.
Notice that in all cases, the transition from downflow to
upflow is sharp, while the downturn of the gas
is much smoother.

\subsection{Two Arm Unmagnetized and Four Arm Cases.}

\begin{figure}
\epsscale{1.0}
\plotone{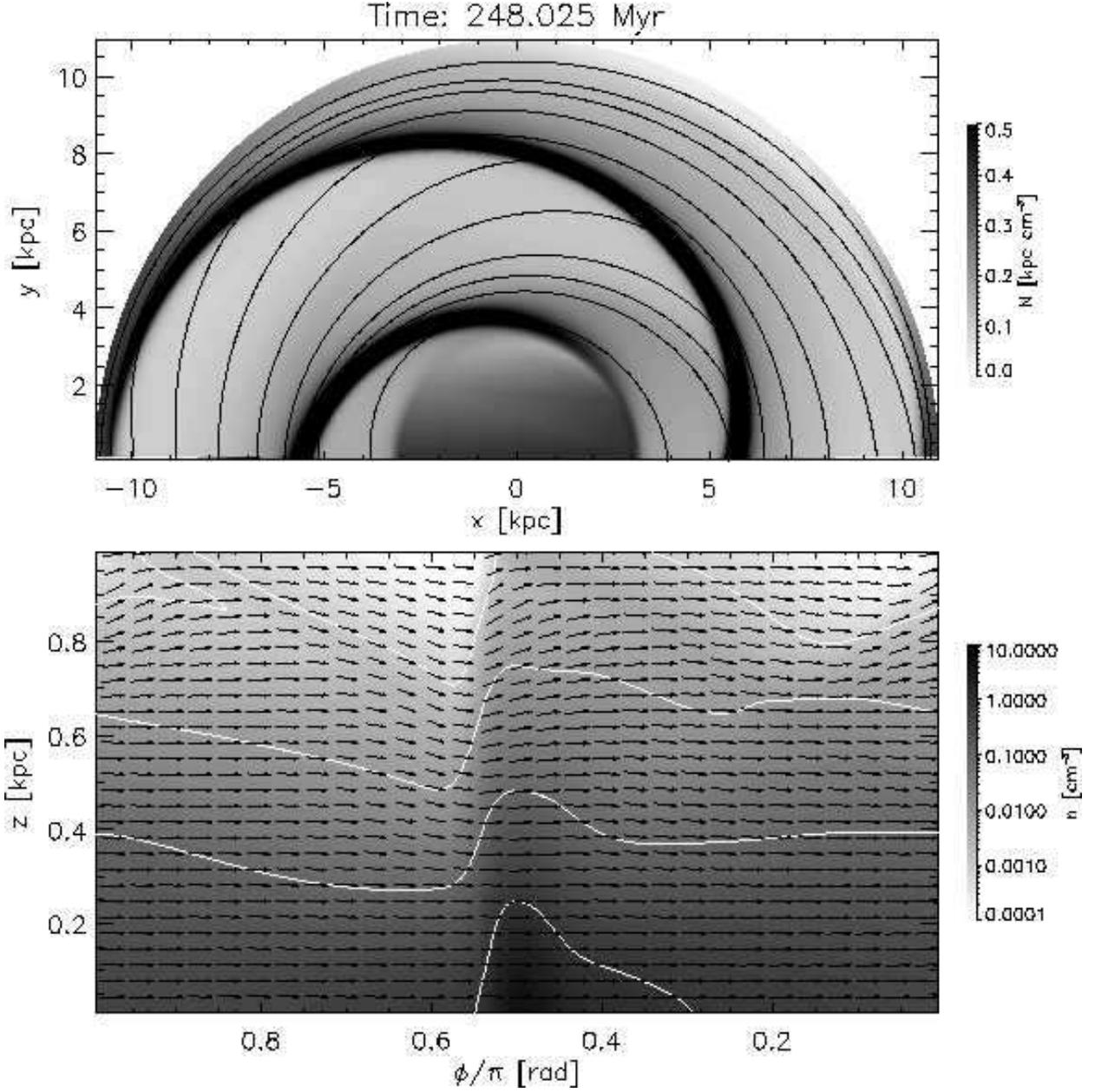}
\caption[]{
Same as Figure \ref{fig3}, but for the \nonmag~ case.
The temperature
in this case is increased to $2.5 \times 10^{4} \degK$ in order
to have a similar equilibrium density distribution without the
magnetic field of the previous case.
The arm shock and density ridge are more vertical
and closer together than in the
magnetized case, and the flow up and over the arms is less extreme.
There is a weak shock at the upper right corner of the lower
panel, roughly from $(\phi,z)= (.2,.8)$ through (0,.9),
where the downflow bounces, but on the whole, the flow
is much less structured, less like a hydraulic jump.
}
\label{fig5}
\end{figure}

\begin{figure}
\plotone{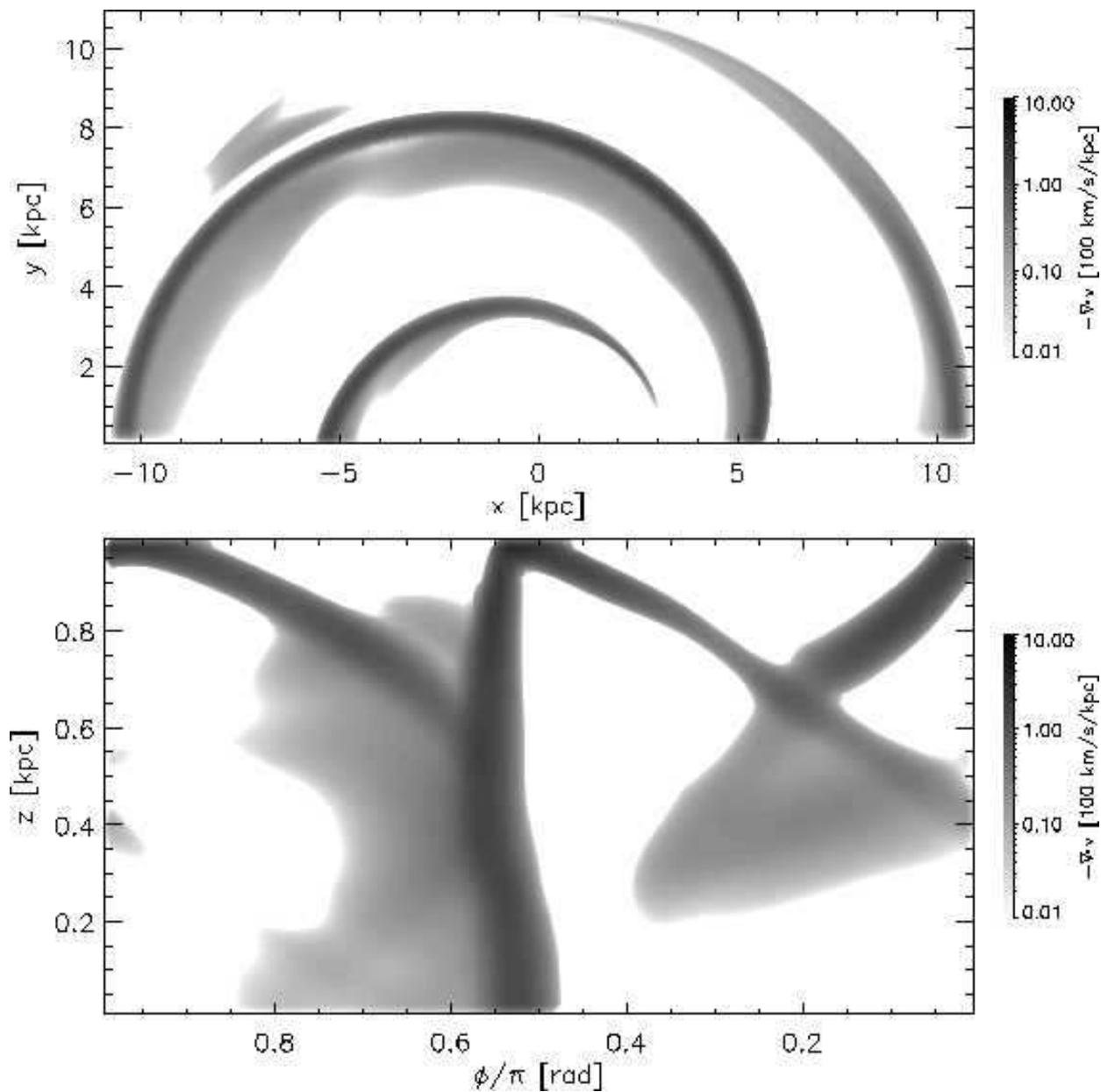}
\caption[]{
Divergence of the velocity for the \nonmag~ case.
Again, only those places with negative divergence are presented.
Notice the presence of an important vertical shock just upstream
from the gaseous arm.
High-$z$ slanted shocks behind the arm like those
observed in the magnetic
case are also present here, but are weaker and may involve
significant interaction with the closed upper boundary.
}
\label{fig9a}
\end{figure}

As a comparision, Figures \ref{fig5} and \ref{fig9a}
show the \nonmag~ case at 248 Myr.
Here, the temperature is higher in order to have a similar
vertical density distribution.
In the upper panel of Figure \ref{fig5},
we again observe the crowding of the
midplane velocity field into the arms, as in Figure \ref{fig3}.
The vertical structure now presents a strong shock at the leading
edge of a nearly vertical gaseous arm and a much smoother density
distribution behind it, which makes the gaseous arms
fuzzier on the downstream side.
There is vertical velocity structure, similar to the previous
case, but with lower magnitude and only in the upper half
of the grid.

\begin{figure}
\plotone{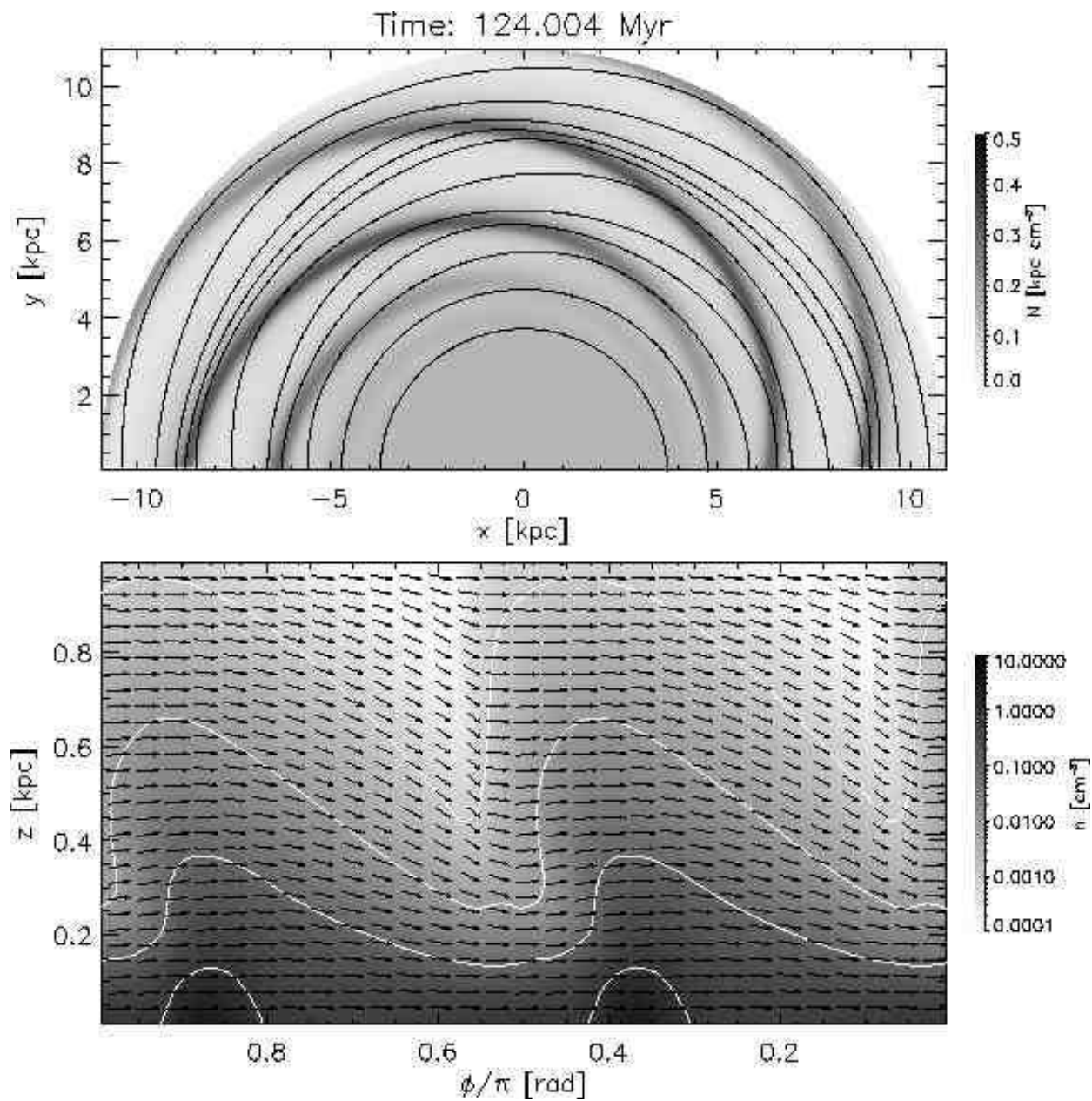}
\caption[]{
The same as Figure \ref{fig3}, but for the \fourarm~ case
at a time of 124 Myr.
The gas, again, proceeds from left to right through the pattern.
The earlier time was chosen so that the number of interactions
with the arm potential would be similar, i. e. the flow is at
a similar level of maturity.
The flow in the reduced interarm space is considerably
less complicated than that of the two arm case, apparently with
insufficient time for an extra bounce.
The column density in the arms is also lower because the
concentration is so high that it matters that the material
is being shared among four arms rather than two.
Notice the substantial downflow preceeding the arm shock and
the subsequent almost imperceptible upflow.
}
\label{fig6}
\end{figure}

At 124 Myr, the four arm case \fourarm~
in Figure \ref{fig6} looks very similar to our standard case.
Above 150 pc, strong downflows before the arms
change suddenly to upflow at the arm position.
In general, the column density is always smaller, since the same
amount of mass is being distributed over 4 arms, leading
to smaller arm to interarm contrast.
This case is identical to the one presented in Figure \ref{fig2},
except for having 4 spiral arms perturbing the
potential\footnote{As discussed in \citet{cox02}, the scale
height for the potential perturbation
also depends on the number of arms involved.
See also \citet{mar02}.} and that the grid covers only a quarter
of the disk.
We do not get interarm structures probably because there is
not enough room between the arms for it.

Another difference between the four and two arm
cases is that, for a given depth,
the potential is steeper when four arms are present.
\citet{mar02} developed a self-consistent model of the
galactic spiral arms which resulted in
narrower arms with a flat interarm region.
They performed 1D MHD simulations with the usual sinusoidal
potential and their modified perturbation, and found differences
in the gaseous structures generated.
The full 3D hydrodynamical effects of the details of the
implementation of the perturbation potential
will require further investigation.

\begin{figure}
\epsscale{0.25}
\plotone{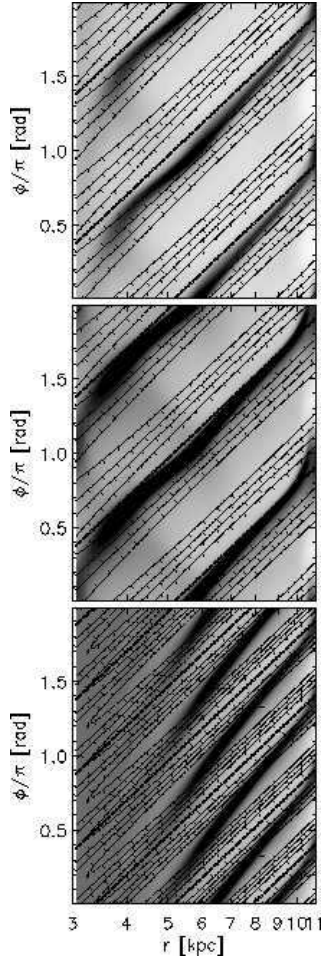}
\caption[]{
Comparison of the density structures and the spiral perturbation
in $\log r - \phi$ space.
In the three figures, the gray scales are column density,
while the solid lines denote the spiral perturbation.
Dotted lines show spirals with a pitch angle of $15 \deg$
and run through the potential minima.
In our trailing spiral model, the gas flows down from the top.
From the top, panels correspond to the \twoarm, \nonmag~ and
\fourarm~ cases.
Our potential perturbation is full strength only outward from
5 kpc.
Notice that in the two arm cases
the density concentrations appear downstream from
the potential well, angling toward it with increasing $r$.
The effect is so strong in the four arm case that the
gaseous arm actually crosses the potential maximum (the stellar
interarm).
In all the cases the gas generates a spiral
slightly tighter than that of the imposed potential.
}
\label{fig7}
\end{figure}

A nice way to examine the phase between gaseous arms and
the perturbing potential is by plotting column density
in $\phi-\log(r)$ space, so that a logarithmic spiral
appears as straight lines.
Such plot is presented in Figure \ref{fig7}.
Solid lines show contours of the underlying potential perturbation
at the midplane,
with the tick marks indicating the downhill direction.
Dotted lines follow logarithmic spirals with pitch angle
equal to the perturbation ($15 \deg$),
along the perturbation minima.
Gray scale indicates column density in arbitrary units.
As our model is of trailing spirals,
the gas flows down from the top.
In the \twoarm~ (upper panel), and \nonmag~ (middle panel)
cases, the gaseous arms are
slightly downstream from the potential minimum, by
a gradually varying amount.
The \fourarm~ case (bottom panel), on the other hand,
follows a tighter spiral with a pitch angle of about $12 \deg$.
Also, the gaseous arms do not extend as far into the inner radii.
If this difference in pitch turns out to be a robust
feature of future simulations, it would lead to differences in the
characteristics of the
spiral structure of galaxies when comparing
observations of Pop II with \ion{H}{1} or Pop I tracers
\citep[for example]{dri00}.

\subsection{Cyclic Variation.}

\begin{figure}
\epsscale{1.0}
\plotone{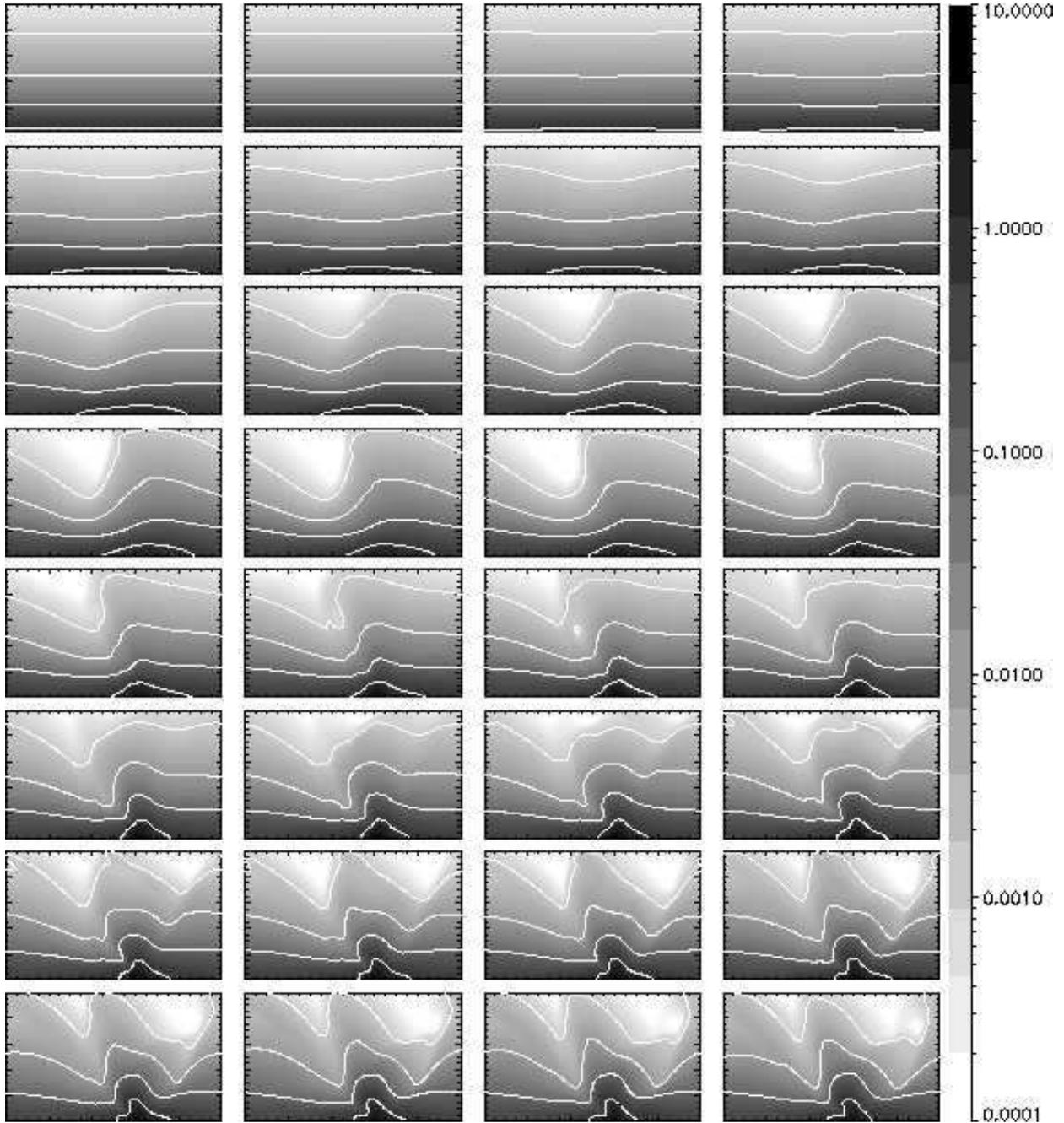}
\caption[]{
Time
sequence of density slices for the \twoarm~ case.
Each panel corresponds to the lower panel of Figure \ref{fig2},
and there are 8 Myr between each slice.
At high $z$ there is some indication of a
``breaking-wave like'' structure
with a period of around 80 to 100 Myr.
The period and structural details likely depend on the height
(1 kpc) and type of the upper boundary (closed).
}
\label{fig10}
\end{figure}

The runs presented here do not yet correspond to an asymptotic
state,
but it appears that they already have some cyclic behavior in
the their evolution. Figure \ref{fig10} presents the time sequence
for the evolution of the \twoarm~ case. Each panel corresponds
to the lower panel in Figure \ref{fig2}, starting at t=0
with 8 Myr spacing.
A high density structure above the arm is fully formed and
leans upstream at $\approx 120 \Myr$.
It then contracts back as material from the tip falls
down ($t \approx 168 \Myr$), only to rise again from
behind the arm, as in a breaking wave.
Although it is hard to see in this Figure, such behavior is
evident in an animation.
The interarm structure (at the left side of the plots) also
shows such cycles, with approximately the same phase.
The reader must keep in mind that the gas making
these structures is constantly moving around the galaxy and,
therefore, these motions are really density waves on top
of the galactic rotation.
These cycles show a period of approximately 80 to 100 Myr.
The \fourarm~ case presents a similar behavior, with a
slightly longer period.
The \nonmag~ case also presents such
a cyclic variation, although it is less prominent.
In all cases, the behavior is superposed on secular evolution
in the arm structures.

\subsection{Later Stages.}

\begin{figure}
\plotone{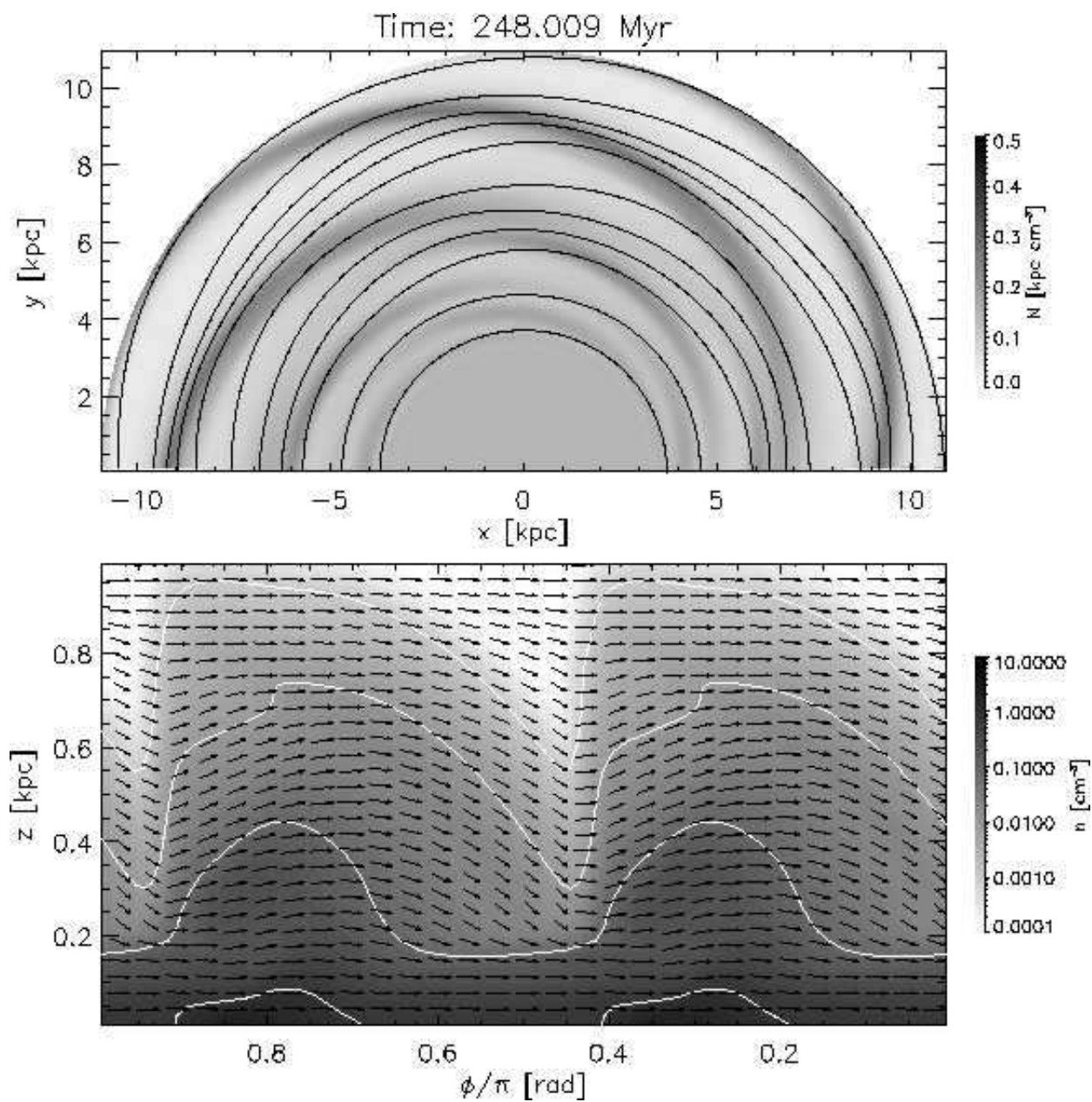}
\caption[]{
\fourarm~ case at 248 Myr. Notice the formation of feathered
arms inside 7 kpc.
The lower panel (at $r=8 \kpc$) shows some
qualitative changes from Figure \ref{fig6}.
The density contours drop more abruptly on the downstream side,
the vertical velocities are more pronounced, and the
midplane density and total column density distributions
are wider and more diffuse.
}
\label{fig13}
\end{figure}

\begin{figure}
\plotone{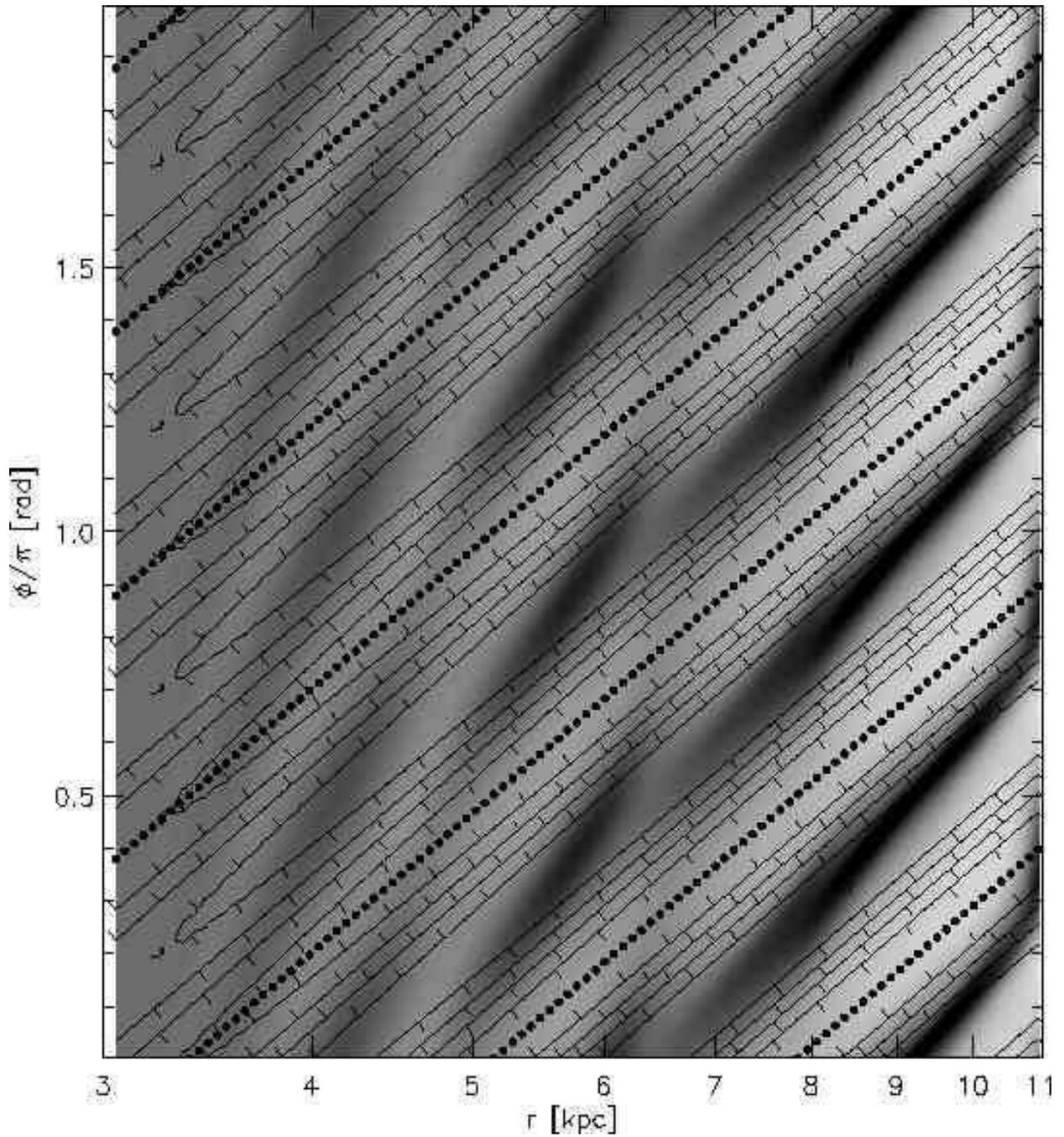}
\caption[]{
Same as Figure \ref{fig7} but for the \fourarm~ case at 248 Myr.
Notice the formation of feathers inside $r=7 \kpc$.
The gaseous arms have locally shallower pitch than the perturbation
but manage, by feathering, to achieve close to the same
average pitch.
The feathers lie along the interarm of the perturbing
stellar arms.
}
\label{fig14}
\end{figure}

Figure \ref{fig13} show the \fourarm~ case at 248 Myr.
At this time, the gas has encountered the spiral arms
twice as many times as in Figure \ref{fig6}.
Fragmented arms are present inside $r=7 \kpc$, while
grand design arms are still present further out.
A $\phi \mbox{~vs.~} \log (r)$ plot (Figure \ref{fig14})
shows the gaseous structures actually distributed along
the potential maxima, instead of the minima.
In the outer edges, the gaseous arms return to their previous
position just downstream of the minimum.
A time sequence of this case show that the gaseous arms drift
downstream and stabilize at the potential crest.
This evolution proceeds from the inner radii out.
After a small section of the arms has
drifted downstream, the arm breaks and the outer tip
remains anchored at its original position.
The result is that each individual
section follows a tighter spiral,
and the locus of all the sections, as a set, follow  nearly the
same spiral as the pertubation potential.
Frequently, observations of the spiral structure in
galaxies, even grand design spirals, show this type
of feathering, in gaseous or Pop I tracers.

\begin{figure}
\plotone{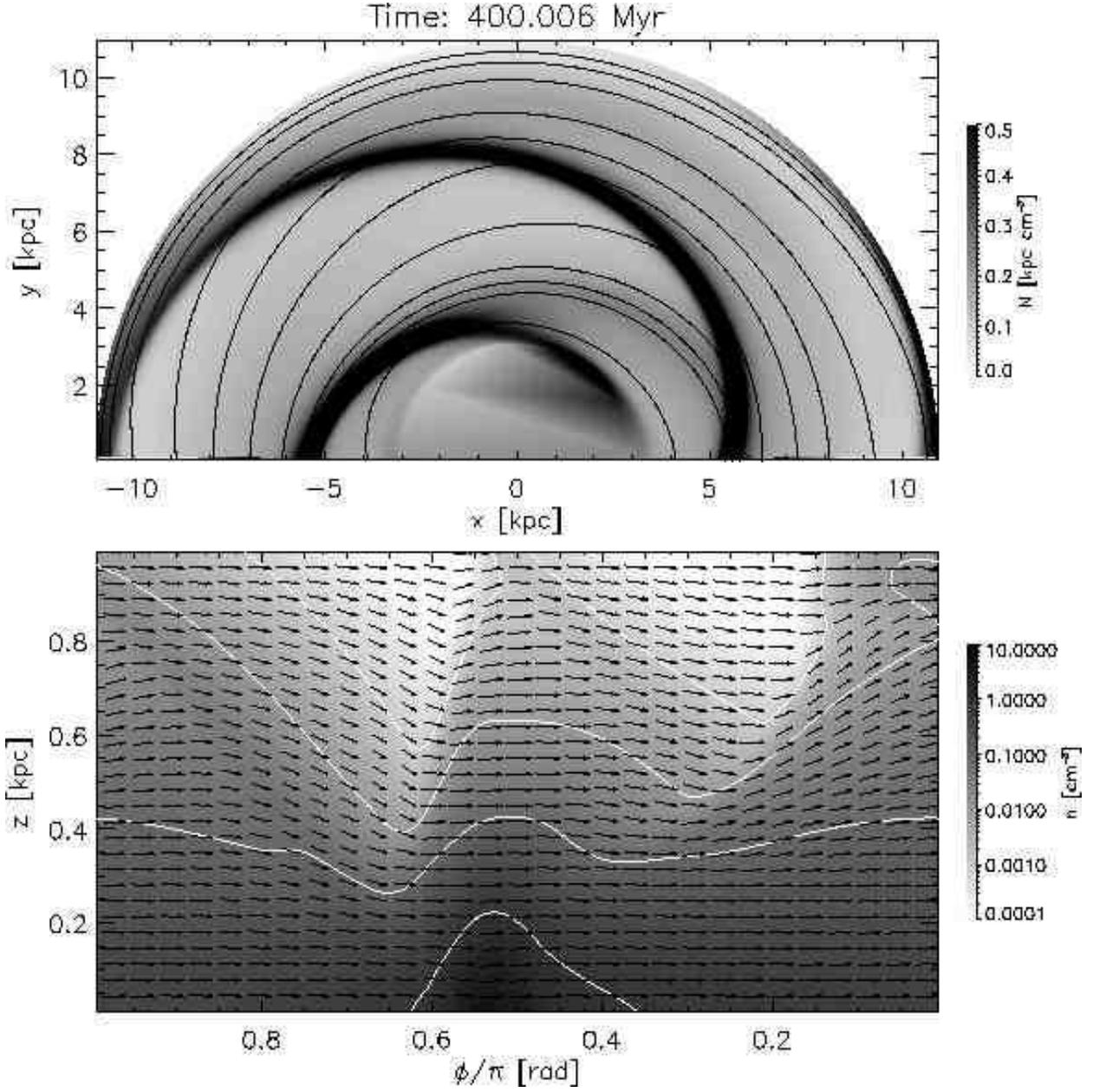}
\caption[]{
\nonmag~ case at 400 Myr.
Notice in the upper panel
the formation of an interarm bridge in the inner galaxy,
stretching between $r=4$ and 5 kpc.
This is the regime in which the amplitude of the
arm perturbation forces are increasing from zero to their
full values.
As with the \fourarm~ case, the arm density structure has become
somewhat more diffuse and the vertical velocity more
pronounced, particularly in the interarm region
($\phi/\pi \approx 0.1$).
}
\label{fig15}
\end{figure}

After 400 Myr, the \nonmag~ case also starts
developing an interarm structure at small radii
(Figure \ref{fig15}), in this case a bridge between arms.
As this entire structure is interior to 5 kpc, the smallest
radius at which our perturbation forces are fully activated,
this structure may be an artifact.
The potential for this formation is already evident from the
velocity field in the upper panel of Figure \ref{fig5}.

\section{SYNTHETIC OBSERVATIONS.}

When we have refined our models and followed them into
asymptotic behavior, we will examine signatures
to compare with observational data.
Here, we present some preliminary examples
of how our simulated galaxy would look to observers
from within, and from the outside.

\begin{figure}
\plotone{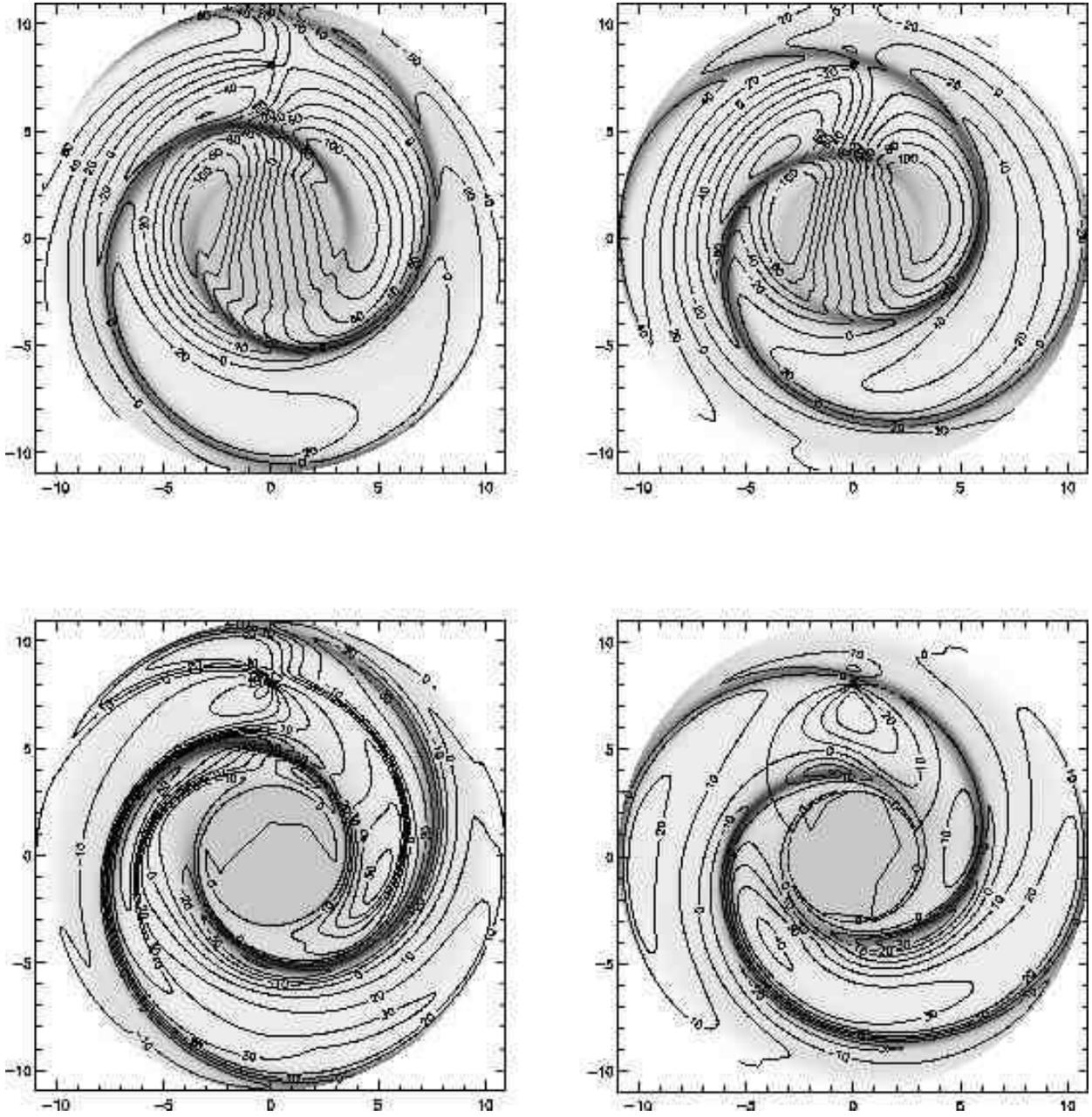}
\caption[]{
The upper panels show the line-of-sight component
of the midplane velocity of the gas
(in km/s) when the observer (at $x=0$, $y=8 \kpc$)
is at a interarm region (left panel)
or near the inner region of an arm (right panel).
The lower panels show the line-of-sight component of the
deviation of those velocities from circular rotation.
Notice that large velocity differences appear even at locations
near the observer.
}
\label{fig19}
\end{figure}

Gas velocities inconsistent with a circular galactic rotation
have been routinely observed (see for example Tripp, Sembach
\& Savage 1993).
In order to study the line-of-sight component of the
non-circular motions in our modeled galaxy,
we assigned a reference position in the midplane of the disk
and calculated the radial velocity from it of all
other midplane locations.
Upper panels on Figure \ref{fig19} show that radial
component, while the lower panels show
the velocity difference between it and
a purely azimuthal rotation.
The two cases are for an observer in an interarm region
and one at the inner edge of a gaseous arm.
The circular velocity considered was the solution to
our hydrostatics equations discussed in Section 2.
The velocity differences frequently exceed $20 \kms$, which
would lead to errors in distance determinations of several kpc,
even near the observer.
Suggestion of problems in the determination of
kinematic distances to pulsars have been found by \citet{gom01}.
After tuning the input parameters in our simulation to a more
realistic picture of the Milky Way, we
will examine the degree to which
those distance inconsistencies can be accounted for
by considering non-circular gas motions.

\begin{figure}
\plotone{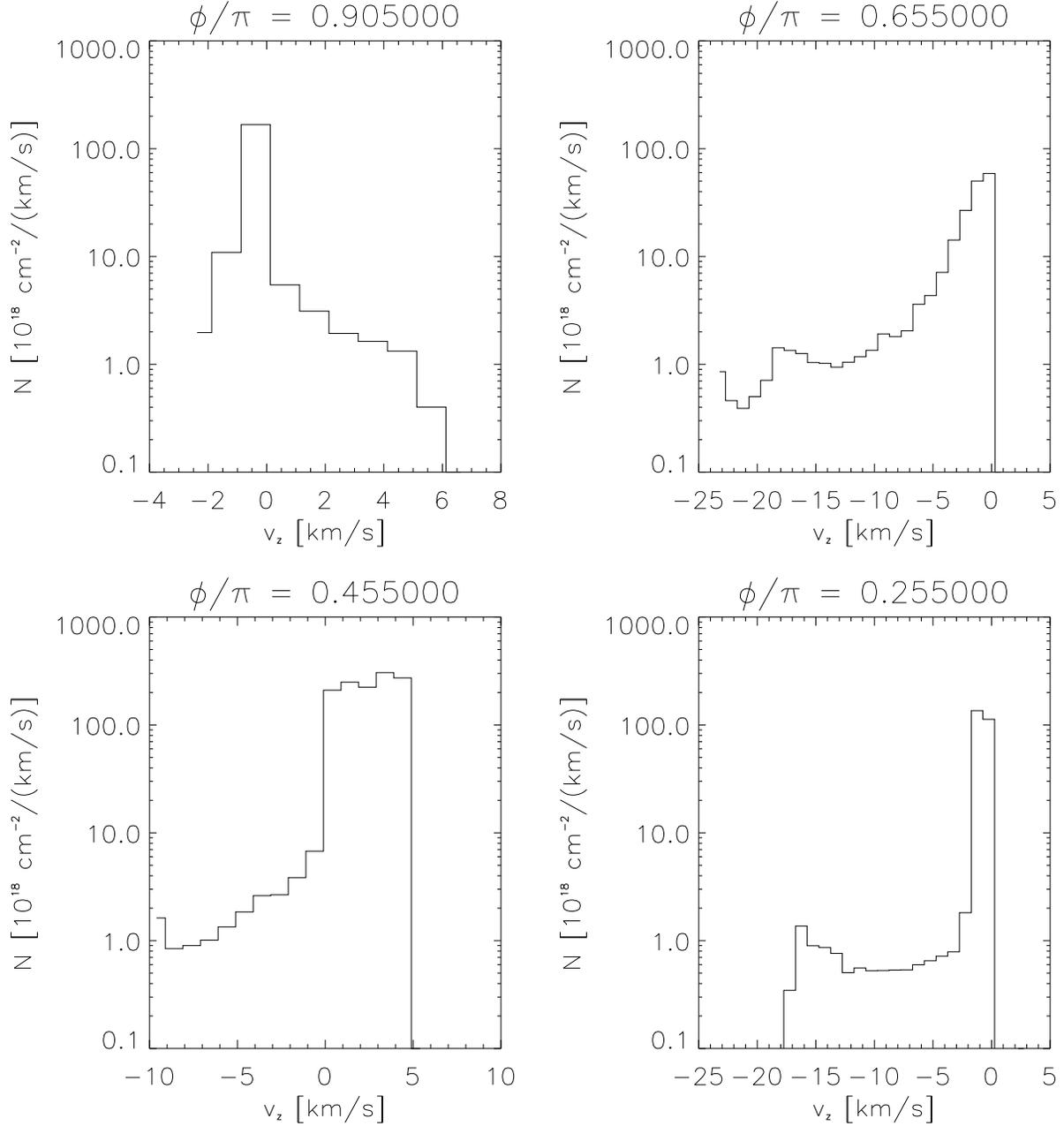}
\caption[]{
Simulated spectra for the \twoarm~ case at $r=8 \kpc$.
The panels are picked trying to catch important features
from Figure \ref{fig3}, namely an interarm region,
at and around the gaseous arm.
Even though most of the material sits near the plane, it is
possible to obtain secondary peaks in velocity space that
do not correspond to real gas concentrations in the physical
space.
}
\label{fig21}
\end{figure}

Motions in the ISM are also observed in the form of intermediate
velocity gas above the plane.
In the simulations presented here, a large layer of gas falls
behind the spiral arms (and sometimes, between them),
all with similar speed.
So, to an observer situated at the right position, this
gas would appear as a falling velocity feature,
a ``cloud,'' even without a localized density enhancement.
Figure \ref{fig21} shows this situation.
After choosing a particular position in radius and
azimuth, we interpolated the vertical distribution to a finer
grid in order to generate a smooth artificial spectrum.
All the spectra in the Figure correspond to $r=8 \kpc$.
Comparison with Figure \ref{fig3} shows that the four
azimuthal positions correspond to interarm, just before
the arm, at the midplane density peak,
and the downflow region after the arm.
The two left hand panels show the characteristic upflow
extending to about $5 \kms$, while the two right hand
panels (and to a lesser extent, the lower left one)
present an extended wing towards larger
negative velocities, together with a small peak at about
$-18 \kms$.
As seen in Figure \ref{fig8},
the asymmetry in the velocity distribution appears because the
upflow happens in a narrower region of higher density and lower
velocity than the downflow.
Therefore,
just by chance, it is easier to pick a region in which the gas
seems to be falling than one with upflowing gas, and the
velocities are likely higher as well.

\begin{figure}
\plotone{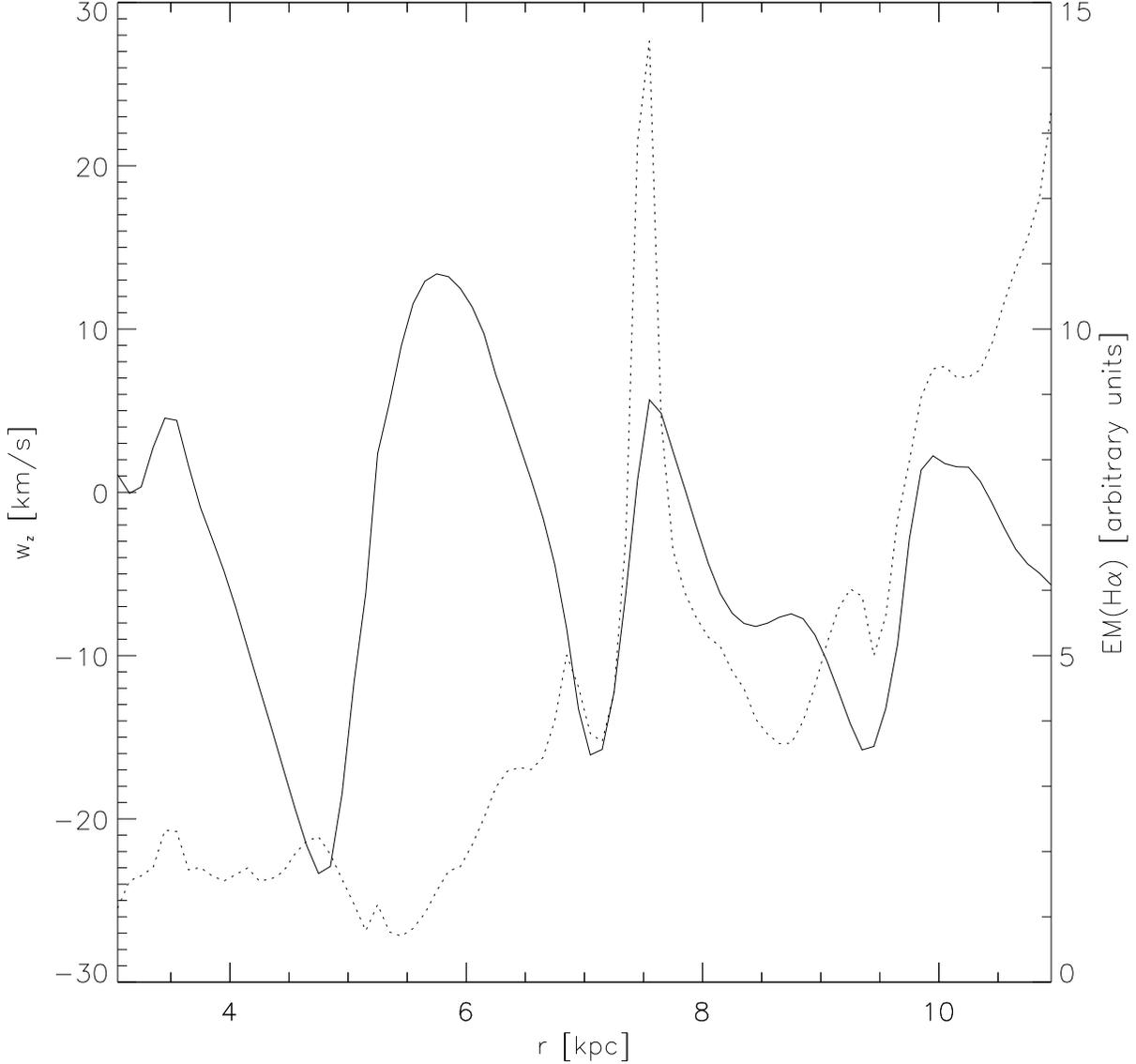}
\caption[]{
Synthetic H$\alpha$ observation of the model along
$\phi=\pi/2$.
This Figure is to be compared to Figure 2 in \citet{alf01}.
The solid line is the vertical velocity weighted with
the square of the density for those grid-zones with
$n < 0.01 \cm^{-3}$.
It is similar in every feature to the vertical
velocity at $z=0.31 \kpc$ in Figure \ref{fig8}.
The dotted line is the emission measure for the same positions.
As the gas approaches the gaseous arm at 8.5 kpc, the flow
rises to pass above it and then falls behind.
Notice that the peak of the EM is upstream from the
higher density gaseous arm in the midplane due to the forward
lean of the arm at high $z$.
}
\label{fig20}
\end{figure}

Our third  example shows the appearance of our galaxy model
from outside, in velocity resolved spectra of lines
originating in the low density material well off the plane.
Vertical motion of the gas have been found in
H$\alpha$ observations of NGC 5427 by \citet{alf01}.
These motions are consistent with what would be expected by
corrugation in the velocity field
of the gaseous disk induced by a hydraulic jump
around the spiral arms.
Our simulations do not include the ionization structure of the
gas (and could not without including the ionizing agents)
but it is reasonable to expect that the lowest density regions
will be ionized and that their behavior will approximate that of
the warm ionized component of the interstellar medium
and have a direct relation with H$\alpha$ observations.
With this in mind, we simulated the study done by \citet{alf01}
in Figure \ref{fig20} using our \twoarm~ case.
The continous line is the integrated vertical velocity
weighted by the square of the density,

\begin{equation}
w_z(r,\phi)=\frac{\sum_z v_z(r,\phi,z) \times [n(r,\phi,z) f(n)]^2}
                 {\sum_z [n(r,\phi,z) f(n)]^2},
\end{equation}

\noindent
where $f(n) = 1 $ if $n \leq n_{lim}$,
$f(n) = (n/n_{lim})^{-2.17}$ otherwise,
and $n_{lim}= 0.01 \cm^{-3}$.
The dotted line show the emission measure for these
same grid points, defined as

\begin{equation}
EM=\sum_z [n(r,\phi,z) f(n)]^2,
\end{equation}

\noindent
in arbitrary units.
The midplane gaseous arm is at $r=8.5 \kpc$.
As observed by \citet{alf01}, the gas moves up as it approaches
the arm, falling behind it.
A distinctive feature is that the EM peak is upstream from
the gaseous arm at the midplane.
Notice that the observations of NGC 5427
are for a region outside corotation, and therefore, the gas
there moves from the convex to the concave sides of the gaseous
arm, in the opposite direction from the gas in our simulation.
In addition, the approximate $30 \deg$
inclination angle for NGC 5427
should also show the imprint of radial streaming motions in the
plane of the galaxy in this kind of observation.

Comparison of our Figures \ref{fig3} and \ref{fig20}
with Figure 2 in \citet{alf01}
shows that the H$\alpha$ profile of a spiral arm is more
ragged than the model profile in higher density gas.
While the column density
plot in our Figure \ref{fig3} is a smooth spiral,
the H$\alpha$ emission measure in Figure \ref{fig20}
has many wiggles.

\section{CONCLUSIONS AND FUTURE WORK.}

In this work, we present our early results in the 3D MHD modeling of
the large scale interaction of the ISM with a spiral potential.
The presence of a thicker, more pressurized gaseous disk,
together with the extra freedom the gas has in 3D simulations,
allows the generation of density and velocity structures that
previous work failed to reveal.

We confirmed and extended the work by MC, in which large scale
vertical motions of the gas are an intrinsic feature of the response
to the spiral perturbation. 
The downflow occurs along a much broader region and at higher
velocities than the upflow.
The falling gas can have large regions with
a very similar vertical component,
which translates into velocity crowding.
In the present models,
this gas appears as peaks at about $20 \kms$
in spectra taken directly up from the midplane.
These motions are accompanied with rapid flow above the arms
and similar ``up and over'' motions in the radial direction.
So far, there are some hints that such motions might be occuring
in NGC 5427 \citep{alf01}.

We also found significant differences in the midplane
line-of-sight velocity distribution as compared with a
purely circular rotation model.
We think that the presence of streaming motions generated by
the spiral arms must be considered when estimating the distance
to elements of the ISM using their velocity as reference.
In the future, when we obtain a more realistic model
for the Milky Way, we may be able to provide a
reasonable recipe for translating
radial velocities and galactic longitude data to distance in a more
reliable way.

Our models have a number of numerical simplifications
(low resolution, closed boundaries, short run times)
and omission of physical processes (heating and cooling
of the gas, cosmic rays, self gravity, ionization, star
formation or associated energy injection).
Improvement on the run times and resolution will allow us
to follow the structures to maturity, better examine cyclic
features, explore substructure formation such as feathers,
bridging and gaseous interarms, follow the magnetic field
energy density and geometry to saturation, and to explore
radial migrations of material and angular momentum. Addition
of a more realistic equation of state to represent
heating and cooling of
the gas will allow the formation of truly dense regions.
The interaction between magnetized flow and these regions
may qualitatively alter the general arm structure, the velocity
field and the complexity of the magnetic field configuration.

Our results show that failure to consider high $z$
and non-circular motions of the
ISM associated just with the response to the spiral potential
can easily lead to confusion when interpreting
observational results.
The study of the gaseous structure of the Milky Way and
other galaxies require the consideration of three dimensional
effects and a more realistic model of the nature of the ISM and
its interaction with other dynamical elements of the system.

\acknowledgements

We thank R. Benjamin, M. Martos and M. Bershady for useful
comments and suggestions, to the NASA Astrophysics Theory
Program for financial support under the grant NAG 5-8417,
and to M\'exico's Consejo Nacional de Ciencia y Tecnolog\'{\i}a
for support to G. C. G.

\clearpage

{}

\clearpage

\clearpage

\clearpage

\clearpage

\clearpage

\clearpage

\clearpage

\clearpage

\clearpage

\clearpage

\clearpage

\clearpage

\clearpage

\clearpage

\clearpage

\clearpage

\clearpage

\end{document}